\documentclass[9pt,twocolumn,twoside]{osajnl}

\journal{ao} 
\setboolean{shortarticle}{false}
\newcommand{\bfN}{\boldsymbol{N}}
\newcommand{\ie}{~\emph{i.e.}}

\title{Partial-Field Illumination Ophthalmoscope: improving the contrast of a camera-based retinal imager}

\author[1,$\dag$]{Léa Krafft}
\author[2,$\dag$]{Elena Gofas-Salas}
\author[1,2]{Yann Lai-Tim}
\author[2,3]{Michel Paques}
\author[1]{Laurent Mugnier}
\author[4]{Olivier Thouvenin}
\author[1,4]{Pedro Mecê}
\author[1,*]{Serge Meimon}

\affil[1]{DOTA, ONERA, Université Paris Saclay F-91123 Palaiseau, France}
\affil[2]{CIC 1423, INSERM, Quinze-Vingts Hospital, Paris, France}
\affil[3]{Institut de la Vision, 17 rue Moreau, Sorbonne Universités, UPMC Univ Paris 06, INSERM, CNRS, 75012 Paris, France}
\affil[4]{Institut Langevin, ESPCI Paris, Université PSL, CNRS, 75005 Paris, France}
\affil[$\dag$]{Co-first authors with equal contribution}

\affil[*]{Corresponding author: serge.meimon@onera.fr}

\begin{abstract}
Effective and accurate in-vivo diagnosis of retinal pathologies requires high performance imaging devices, combining a large field of view and the ability to discriminate the ballistic signal from the diffuse background in order to provide a highly contrasted image of the retinal structures. Here, we have implemented the Partial-Field Illumination Ophthalmoscope, a patterned illumination modality, integrated on a high pixel rate adaptive optics full-field microscope. This non-invasive technique enables us to mitigate the low signal-to-noise ratio, intrinsic of full-field ophthalmoscopes, by partially illuminating the retina with complementary patterns to reconstruct a wide field image. This new modality provides an image contrast spanning from the full-field to the confocal contrast, depending on the pattern size. As a result, it offers various trade-offs in terms of contrast and acquisition speed, guiding the users towards the most efficient system for a particular clinical application.
\end{abstract}

\begin{document}

\maketitle

\section{Introduction}
In order to provide a better understanding and early diagnosis of age-related macular degeneration (AMD), it is critical to be able to follow the retinal degeneration, and in particular the loss of photoreceptor cells \cite{curcio_photoreceptor_1996}.
However, this task requires high resolution imaging systems, able to provide highly contrasted photoreceptor maps with a wide field of view. The need for distortion-free images is also essential, particularly for longitudinal studies, when distinguishing the morphological characteristics of the retinal structures \cite{ kolb_webvision_1995} and their evolution through time should not be hindered by distortion artifacts coming from the imaging system and the ocular motion \cite{salmon_automated_2017,cooper_effects_2016}.
 Therefore, there is a need for highly resolved, distortion-free images of the retina, with a high contrast and a wide field of view. 
 
 To this end, ophthalmoscopes have to mitigate aberrations, ocular motion \cite{jarosz_high_2017, mece_fixational_2018}
 and a high diffuse background originating from scattering effects in the multi-layered retina \cite{burns_adaptive_2019}. Regarding aberrations, Adaptive Optics (AO)  
 has enabled high resolution
images of the photoreceptor layer in both full-field ophthalmoscopes \cite{liang_supernormal_1997} and Scanning Laser Ophthalmoscopes (SLO) \cite{roorda_adaptive_2002}, revealing structures that were hidden otherwise \cite{godara_adaptive_2010, burns_adaptive_2019, paques_adaptive_2018}. 
 
 Regarding ocular motion and background concerns, a trade-off has been made either in favor of motion handling, with full-field ophthalmoscopes or in favor of contrast \cite{gofas-salas_high_2018,liang_supernormal_1997} with the SLO. The full-field ophthalmoscope simultaneously acquires the signals coming from all the positions of its wide illumination field ($4^{\circ}$ typically), with exposure time below 10 milliseconds, which prevents motion-induced distortions. On the other hand, the confocal pinhole of the SLO spatially filters out multiply scattered and out-of-focus light hence improving the quality of the images. However, the full-field ophthalmoscopes remain limited in contrast, while the SLO systems are prone to distortion due to their low pixel rate, which clearly illustrates the complementarity in strengths and weaknesses of these techniques.
 
In this paper, we implemented the Partial-Field Illumination Ophthalmoscope (PFIO), a high pixel rate AO-full-field ophthalmoscope in which the illumination field can be tuned in order to reduce the diffuse background and improve the contrast of photoreceptor cell images. First, we demonstrate that spatial filtering can be achieved in a full-field ophtalmoscope by partially illuminating the retina (by placing a programmable mask conjugated with the retina in front of the full-field illumination source). By setting the camera pixels associated with non-illuminated areas on the acquired frame to zero, a numerical spatial filtering is performed similarly to using a variable confocal pinhole in SLO. We establish that the contrast can be improved proportionally to the square root of the surface of the illuminated area. The full field image can then be obtained by this process (\ie  image acquisition and numerical spatial filtering) for all the complementary patterns required to illuminate the entire field and by combining these successive images. The resulting image therefore has an increased contrast at the cost of reduced acquisition speed.
 
\begin{figure}[!h]
\centering
\includegraphics[trim=10 5 25 5,clip, width=\linewidth]{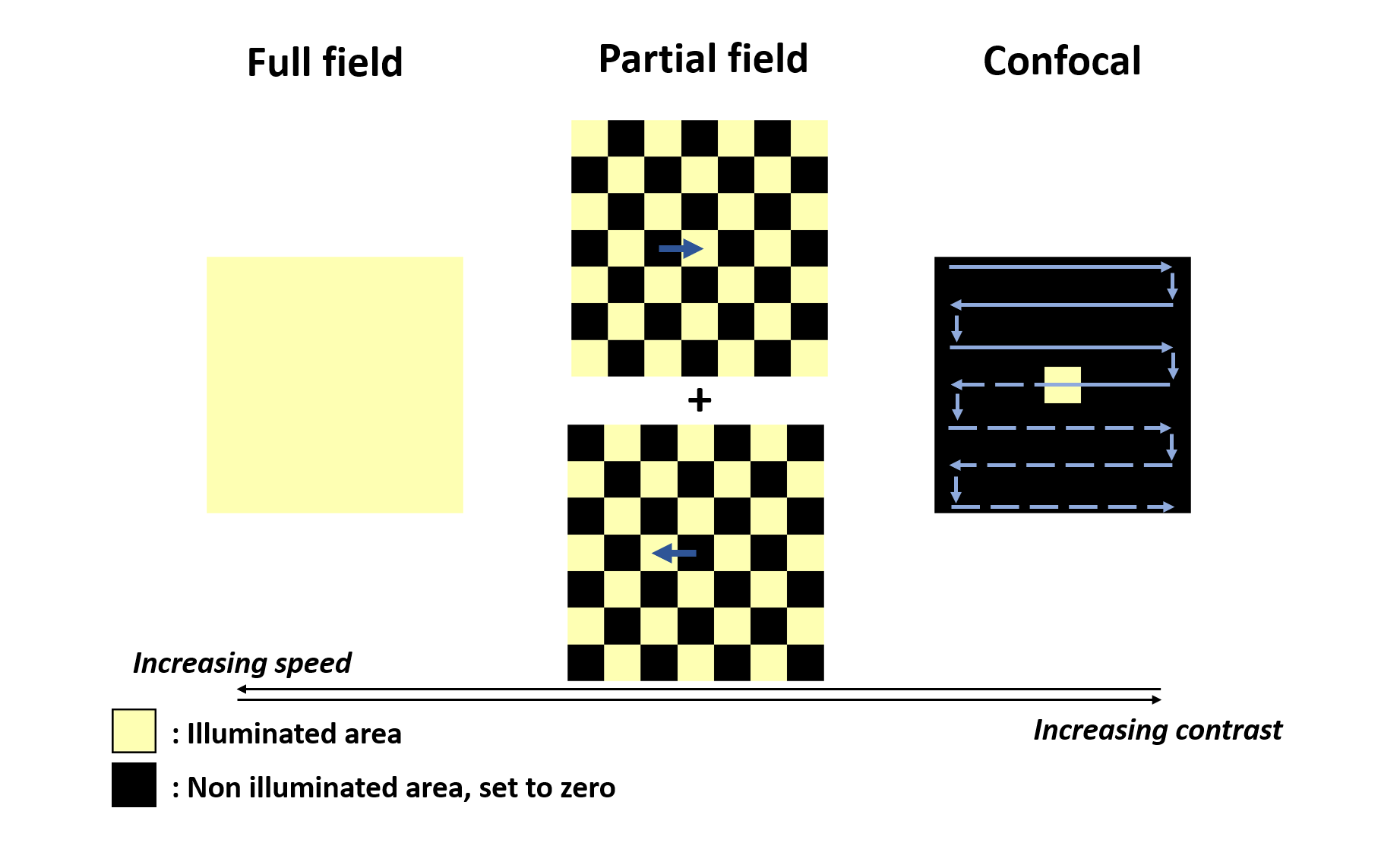}
\caption{Comparing the full-field ophthalmoscope, the partial-field illumination ophthalmoscope and the confocal ophthalmoscope. Schematic drawing of the characteristic patterns utilized in each technique, similar for the illumination and the detection.}
\label{fig:scheme}
\end{figure}

 Our work aims to demonstrate that this partial-field illumination ophthalmoscope can be positioned between full-field and confocal opthalmoscopes. The characteristic illumination patterns of these three ophthalmoscopes are shown in \textbf{Fig.~\ref{fig:scheme}}. For sake of comparison, let us consider the same illumination power in the illuminated area, and the same amount of overall power to compute a complete full-field image. Thus, the acquisition time of the full-field image scales with the number of complementary patterns $\bfN$ used to recontruct the full field image. 
The full-field ophthalmoscope only uses one pattern ($\bfN=1$) to illuminate the entire region of the retina to be imaged. Therefore, it is the fastest system. However, as no spatial filtering is performed, it provides images with the lowest contrast.
In contrast, in the confocal system, the illumination beam diameter corresponds approximately to the Airy disk, so it requires scanning of the region of interest with about one million points ($\bfN\simeq 10^6$) to illuminate the entire field of view for an image 2000*2000 pixels sampled at Nyquist frequency. The acquisition speed is thus the slowest. In the detection path, a physical confocal hole carries out spatial filtering that enables providing the images with the highest contrast. 
Between these systems, the partial-field illumination ophthalmoscope  illuminates the region of interest partially with an intermediate number of complementary checkerboard patterns (e.g. $\bfN=2$ in \textbf{Fig.~\ref{fig:scheme}}). 
As a result, the three ophthalmoscopes  in \textbf{Fig. \ref{fig:scheme}} feature an increasing contrast from left to right, and an increasing speed from right to left.

\section{Method}

\subsection{Setup}
\begin{figure}[b!]
\centering
\includegraphics[width=\linewidth]{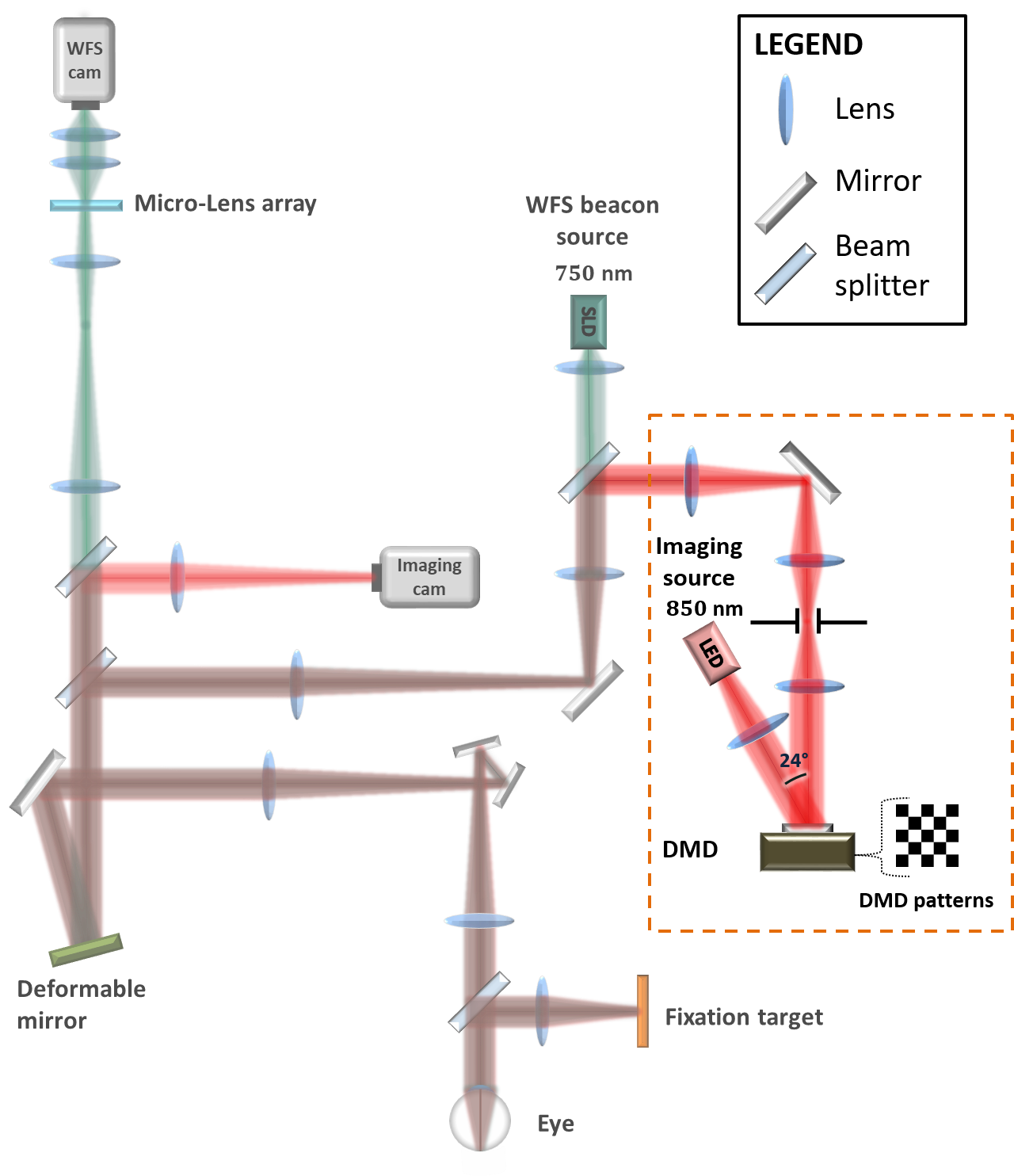}
\caption{Experimental set-up: AO-Full-field optical bench described in \cite{gofas-salas_high_2018}, to which an illumination path, composed with a DMD, has been implemented to project checkerboard patterns on the target layer of the retina.}
\label{fig:setup}
\end{figure}

The experimental setup has been implemented on the AO-full-field ophthalmoscope, built at the Paris Quinze-Vingt National Ophthalmology Hospital, described in \cite{gofas-salas_high_2018} and shown in \textbf{Fig.~\ref{fig:setup}}. We included a new illumination path which consists of a homogeneous incoherent illumination at 850 nm and a spatial light modulator that generates and switches the checkerboard patterns. The patterns are aberration-free, as the light coming from the illumination path reaches the deformable mirror of the AO system of the AO-full-field ophthalmoscope before reaching the eye. Then, the backscattered light coming from the eye encounters the deformable mirror again before reaching the imaging camera. To perform  the various checkerboard patterns, we use an off-the-shelf digital micro-mirror device (DMD) (DLP650LNIR, Texas Instrument) for its high switching speed (up to 12.5 kHz) and its versatility in terms of pattern designs. The array of $1280\times800$ digital micro-mirrors is conjugated to the target layer of the retina and each micro-mirror of $10.8~\mu$m pitch is switched to position “ON” or “OFF” to shape the illumination into the selected pattern to be projected. The mirrors in the “ON” position reflect the incoming light towards the optical path, while the mirrors in the "OFF" position reflect the light out of it. Each pattern, leading to illuminated and non-illuminated  areas on the retina, has been uploaded to the controller of the DMD, manipulated through a computer.
In this configuration, the mirrors, which are used as switches that shape the illumination, are comparable to multiplexed pinholes. Therefore the DMD, which switches from one pattern to the other, behaves similarly to a Nipkow disk used in spinning disk microscopy
\cite{stehbens_imaging_2012, enoki_single-cell_2012}.

\subsection{Pattern projection}
 Patterns were created with Interactive Data Language (Exelis, USA). We used four sets of complementary checkerboard designs, with the same dimension of squares. Each checkerboard design corresponds to $\bfN$ complementary patterns that illuminate a ratio of one out of $\bfN$ squares, with $\bfN = 2, 4, 9, 25$. We aim at a real-time enhancement of the contrast of photoreceptors, namely, a minimal processing of the raw video frames. To do so, images with a field of view of $4^{\circ}\times4^{\circ}$ were acquired at 90 Hz, with an exposure time of 5 milliseconds, while projecting sequentially the complementary patterns on the retinal layer. The checkerboard squares projected were $44.1~\mu$m wide on the retinal plane and the patterns were switched every two frames. The size of the projected squares was chosen to correspond to a whole number of camera pixels.

\subsection{Image acquisition and processing}\label{Image acquisition and processing}
In vivo retinal imaging was performed on a healthy subject, on the partial-field illumination ophthalmoscope. Another imaging sequence was acquired with the standard AO-full-field ophthalmoscope illumination under the same experimental conditions, as a control experiment. As a basis for comparison to typical AO-SLO images, acquisitions of the same region were also performed with the multimodal adaptive optics retinal imager (MAORI) (Physical Sciences, Inc., Andover, MA, USA) at the Quinze Vingts Hospital in Paris, described in \cite{hammer_multimodal_2012}. The participant followed institutional guidelines and adhered to the tenets of the declaration of Helsinki. Informed consent was obtained from the subject after the nature and possible outcomes of the study were explained. The subject was seated in front of the AO-full-field ophthalmoscope and stabilized with a chin and forehead rest and was asked to stare at the fixation target, which is an image of a blue crosshair with a moving dot, enabling us to guide the subject’s line of sight and explore the retina. Image acquisition sessions were performed in standard conditions with neither pupil dilation nor cycloplegia, in a dark room, leading to the largest accessible natural pupil dilation. The total light power entering the eye from the illumination source and the wavefront sensor source are respectively $2000~\mu$W and $2~\mu$W, which is below the power values stipulated by the ocular safety limits, established by ISO standards for group 1 devices.

For each raw image acquired with the partial-field illumination ophthalmoscope, corresponding to $\bfN$ ranging from 2 to 25, ten $32~\mu$m $\times~32~\mu$m regions were selected inside bright areas to compute the Michelson contrast, which corresponds to the ratio $\frac{I_{max}-I_{min}}{I_{max}+I_{min}}$, where $I_{max}$ and $I_{min}$ are the maximum and minimum intensities in each of these regions. We derived the mean and standard deviation of the contrast for each pattern from these values. 

The sequence of raw images, corresponding to N=4, went through processing covering the selection of images (discarding blurred images due for instance to eye motion or intermediate state of the micro-mirrors between two subsequent patterns), the crop of a region of interest of 1024$\times$1024 pixels, the manual registration of the consecutive frames of complementary projected patterns, where individual photoreceptors could be matched in the overlap region, followed finally by the summing of these frames. As a result, $2.54^{\circ}\times2.54^{\circ}$ images (0.75 mm $\times0.75$ mm) of the photoreceptor layer are reconstructed. A residual grid coming from the projected patterns is still visible after reconstruction and was not removed thereafter to allow a fair comparison between the full-field, partial illumination and confocal ophthalmoscopes.

\section{Results and discussion}
\begin{figure*}[ht!]
\centering
\includegraphics[width=\linewidth]{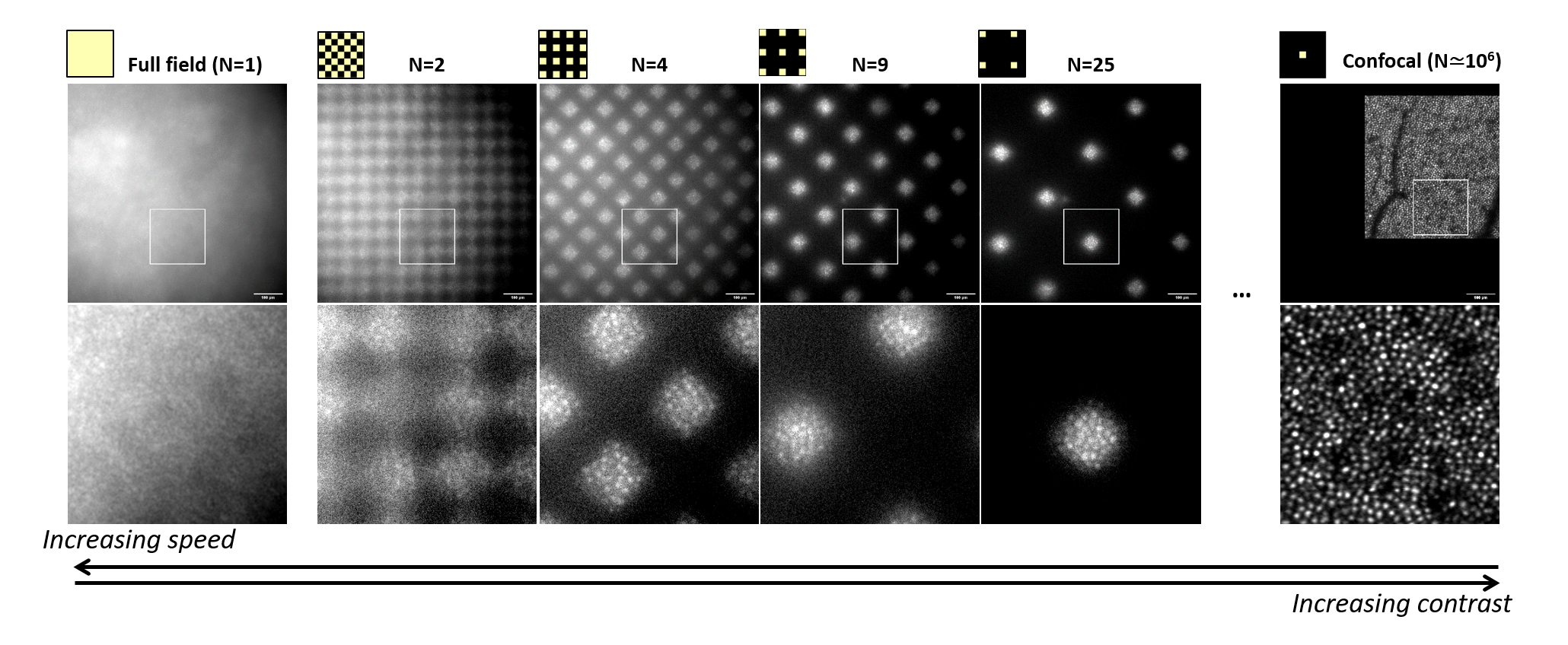}
\caption{First row : Single raw wide field images acquired with the full-field ophthalmoscope (N = 1), partial-field illumination ophthalmoscope: N = 2, 4, 9, 25 and the confocal ophthalmoscope ($\bfN\approx{10^6}$), with $\bfN$ indicating the number of complementary frames necessary for reconstruction of the complete field of view.
Second row: magnified images of $256\times256$ pixels from the first row.}
\label{fig:MCvalues}
\end{figure*}

The checkerboard designs described in the Method section were projected on the same area of the retina, under the same experimental conditions. From these data, we aim at evaluating the contrast of the photoreceptors with different ratios of illuminated areas over the field of view, namely when changing the parameter N.
In \textbf{Fig.~\ref{fig:MCvalues}}, we show the acquired images for $\bfN$ ranging from 2 to 25, over a wide field of view ($1024\times1024$ pixels) and over the same ROI of $256\times256$ pixels. We compare them to both the full-field image (N=1), acquired with the same optical path, and with the confocal image ($\bfN\approx{10^6}$). All images are obtained with the same amount of incident photons. The images are in accordance with our first assessment as we observe a significant enhancement of the imaging contrast as $\bfN$ increases.
The haze is a prominent feature of the full-field image, and is also quite visible in the dark areas of the partial-field illumination ophthalmoscope images. We can see that it is gradually reduced when increasing $\bfN$. As a result, the photoreceptors appear more contrasted and it becomes easier to distinguish them individually. 

\begin{figure}[ht!]
\centering
\includegraphics[width=0.3\textwidth]{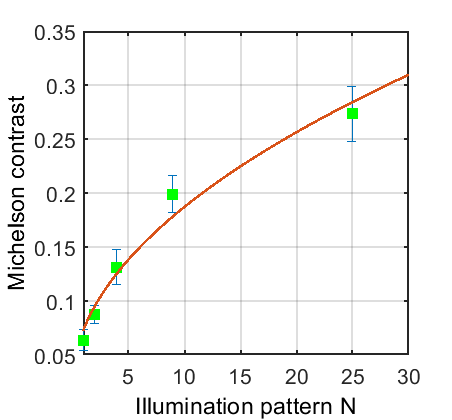}
\caption{Plot of the average and the standard deviation of the Michelson contrast values for each illumination configuration and its fit following a power law proportional to $\sqrt{N}$. For the confocal image ($\bfN\approx{10^6}$), the Michelson contrast was measured at 0.8, which would be obtained with our setup for a value of $\bfN$ between 200 and 1000, according to the fit.}
\label{fig:MCvalues_plot}
\end{figure}

In \textbf{Fig. \ref{fig:MCvalues_plot}}, we aim at quantifying this tendency by plotting the average Michelson contrast, corresponding to each illumination pattern. The bright regions where the contrast was computed were selected at different locations in the field of view to mitigate the influence of a possible inhomogeneity of the background noise.
The Michelson contrast over the illumination pattern $\bfN$ follows a $\sqrt{\bfN}$ fit, as displayed in red. We computed a first degree polynomial fit of the data to the equation model $p(1)\times\sqrt{x}$ in Matlab (using the 'polyfit' function).

The appearance of the photoreceptors in \textbf{Fig. \ref{fig:MCvalues}} and the plot of the Michelson contrast in \textbf{Fig. \ref{fig:MCvalues_plot}} confirm the hypothesis in \textbf{Fig. \ref{fig:scheme}} that the contrast increases with $\bfN$, more precisely as $\sqrt{\bfN}$. Following this trend, a confocal ophthalmoscope with, say, a ratio $\bfN\approx{10^6}$ between the field-of-view and the size of the confocal spot (\ie  one Airy disc in this case), should yield a Michelson contrast $\sqrt{\bfN}\approx 1000$ times higher than the contrast of the full-field ophalmoscope. Our finding is that the contrast gain of AOSLO is less than 1000 times higher then full-field. The experimental Michelson contrast of the confocal image is indeed only 15 to 30 times better, which corresponds to a value of $\bfN$ between 200 and 1000. This means that the contrast improvement with $\bfN$ saturates, or conversely that an image with the same contrast as a confocal image could be obtained with a PFIO requiring 5000 times less illumination patterns than what is currently needed in the confocal system. Therefore, this potential technical improvement could drastically reduce the acquisition time required for the highest imaging contrast and fosters further investigation towards contrast enhancement using the partial-field illumination ophthalmoscope.

\begin{figure*}[!ht]
\centering
\includegraphics[trim=5 1 1 1,clip,width=\linewidth]{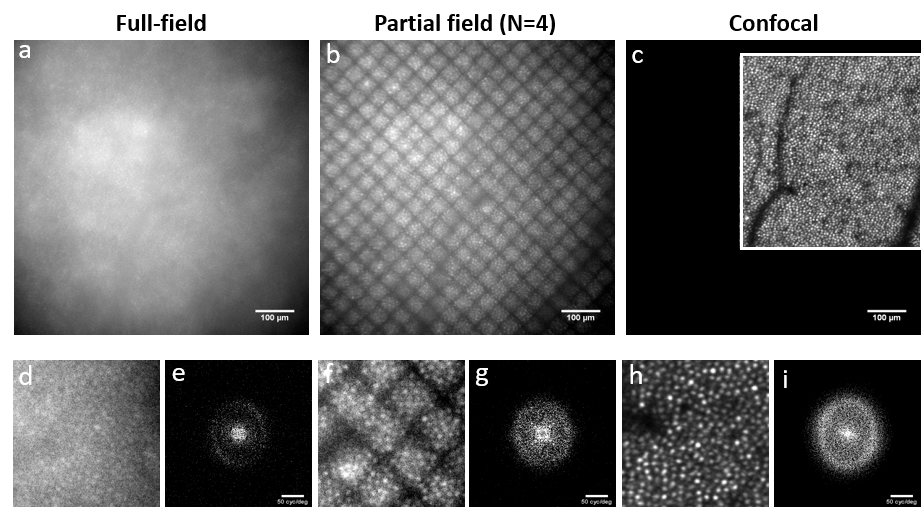}
\caption{Acquired data on AO-full-field ophthalmoscope, partial-field illumination ophthalmoscope and AO-SLO. First row: a) Raw image acquired on the AO-full-field setup with the conventional full field illumination, b) Reconstructed image obtained with the partial-field illumination ophthalmoscope for $\bfN$ = 4 
c) Registered AO-SLO image of the same region. Second row: a magnified image of $200\times200$ pixels ROI of each raw image d), f), h) and their corresponding Fourier transform images, both obtained from an identically located ROI on each image, e), g), i). }
\label{fig:results}
\end{figure*}

Now that we have discussed the theoretical optimal value of $\bfN$ (supported by experimental measurements), let us investigate the practical implementation of PFIO to obtain a complete full-field image with an improved contrast. For $\bfN=4$ for instance,  
we compare the images of the same region, acquired with the full-field ophthalmoscope, the partial-field illumination ophthalmoscope and the confocal ophthalmoscope, under the same experimental conditions, with an equivalent emitted photon flux over the resulting illuminated field. In \textbf{Fig. \ref{fig:results}} - b and f, we show the resulting image obtained with the partial-field illumination ophthalmoscope (see Methods section). The visible residual grid coming from the pattern, which could be removed through processing or by using overlapping complementary patterns, does not hinder the constrast assessment on the image and was kept for a fair comparison between the three systems. While in the full-field image, in \textbf{Fig. \ref{fig:results}} - a and d, the photoreceptors are barely visible in a hazy background, the latter is filtered out in the reconstructed image obtained with the partial-field illumination ophthalmoscope, making photoreceptors appear more defined, brighter and even individually distinguishable. The spatial filtering performance of our technique can also be witnessed in Fourier space, as \textbf{Fig. \ref{fig:results}} - e, g and i particularly shows the computed Fourier transform of the three images. The photoreceptor frequency footprint, or Yellot’s ring \cite{yellott_spectral_1982} appearing gradually brighter from left to right, clearly demonstrates the contrast enhancement of the photoreceptor mosaic.

Each of the four partial-field images used to compute \textbf{Fig. \ref{fig:results}} - b image was obtained with a $\tau_{exp}=5 ms$ exposition (corresponding to the fastest frame rate achievable with our current camera), so that the total exposure time amounts to $\bfN\times \tau = 20 ms$. For higher values of $\bfN$, such an exposure time would be unpractical to efficiently observe the retina, which is why we did not attempted to produce PFIO complete images for values of $\bfN$ higher than 4. In order to accelerate PFIO image delivery rate and get closer to an optimal value of $\bfN$ which we estimate to be between 200 and 1000, a more powerful source and a faster camera could be used (this is indeed the case in confocal imaging, for which the illumination power per surface unit is much higher while the exposure time corresponding to each surface unit is much lower than in our apparatus).  

Taking it one step further, the partial-field illumination ophthalmoscope could spread across an even larger range of imaging techniques. One straightforward application is the use of the partial-field illumination ophthalmoscope approach to carry out dark-field imaging on a camera-based retinal imager \cite{meimon_manipulation_2018, gofas-salas_near_2019, mece_optical_2020, gofas-salas_improvements_2020}. Indeed, by setting the pixels corresponding to illuminated regions to zero, instead of non-illuminated regions, dark-field images could be obtained in a post-processing manner. Although such idea seems simple to put into practice, an important illumination power is necessary, so images from complementary areas can be registered and summed to achieve an entire field-of-view montage. Another interesting application is to couple the partial-field illumination ophthalmoscope with the recently demonstrated structured illumination ophthalmoscope \cite{lai-tim_super-resolution_2020, lai-tim_imagerie_2020}. Indeed, as in the structured-illumination ophthalmoscope the super-resolution is limited by the contrast, using the partial-field illumination ophthalmoscope would allow one to achieve even higher super-resolution, since the achievable super-resolution directly depends on the SNR\cite{lai-tim_jointly_2019}. Finally, the partial-field illumination ophthalmoscope approach could also be a valuable concept to full-field optical coherence tomography (FF-OCT) imaging systems applied for retinal imaging \cite{mece_high-resolution_2020, mece_coherence_2020, scholler_adaptive-glasses_2020}. The use of the partial-field illumination ophthalmoscope, in fact, could reduce cross-talk (in coherent illumination) \cite{auksorius_crosstalk-free_2019}, and/or incoherent light level detected by the 2D camera, contributing to an improvement of SNR and contrast of full-field OCT images \cite{scholler_adaptive-glasses_2020}.

\section{Conclusion}
With the partial-field illumination ophthalmoscope, we propose a new modality that lies between the full-field ophthalmoscope and the confocal ophthalmoscope. By sequentially projecting $\bfN$ complementary checkerboard patterns over the entire field of view and digitally cancelling the non-illuminated pixels, the partial-field illumination ophthalmoscope yields a contrast improvement that scales with $\sqrt{\bfN}$ .The calculation of the Michelson contrast pointed out how the contrast of the confocal system ($\bfN\approx{10^6}$) could be achieved with a value of $\bfN$ several orders of magnitude lower, in other words, at a significantly higher speed. In this way, the partial-field illumination ophthalmoscope holds promise for widening the range of possibility of existing systems.    

\section{Funding}
The research leading to these results has received funding from the Agence Nationale de la Recherche (ANR) EyeWin (grant number ANR-18-CE19-0010), the RHU LIGHT4DEAF (grant number ANR-15-RHUS-0001) and from ONERA's internal project PRF TELEMAC.

\section{Acknowledgments}
The authors want to thank Kate Grieve who provided helpful comments to improve and clarify this manuscript.

\section{Disclosures}
The authors declare no conflicts of interest.

\bibliography{PFIO}

\bibliographyfullrefs{PFIO}

\end{document}